\documentclass{emulateapj}
\newcommand{\rev}[1]{{#1}}
\newcommand{\revs}[1]{{#1}}

\begin{document}

\title{
Rapid Formation of Saturn after Jupiter Completion 
}

\shorttitle{
Rapid Formation of Saturn 
}
\shortauthors{
Kobayashi, Ormel, Ida 
}

\author{Hiroshi Kobayashi\altaffilmark{1,}\altaffilmark{2}, 
Chris W. Ormel\altaffilmark{3,}\altaffilmark{4} and 
Shigeru Ida\altaffilmark{5}
} 

\altaffiltext{1}{
Department of Physics, Nagoya University, Nagoya, Aichi 464-8602, Japan
}
\email{hkobayas@nagoya-u.jp}

\altaffiltext{2}{Astrophysical Institute and University Observatory,
Friedrich Schiller University, Schillergaesschen 2-3, 07745 Jena, Germany}

\altaffiltext{3}{
Astronomy Department, University of California, Berkeley, CA94720, USA
}
\email{ormel@astro.berkeley.edu}
\altaffiltext{4}{Hubble Fellow.}

\altaffiltext{5}{Tokyo Institute of Technology, Ookayama, Meguro-ku, Tokyo 152-8551, Japan}
\email{ida@geo.titech.ac.jp}

\begin{abstract}

We have investigated Saturn's core formation at a radial pressure
 maximum in a protoplanetary disk, which is created by gap
 opening by Jupiter.  A core formed via planetesimal accretion induces
 the fragmentation of surrounding planetesimals, which generally
 inhibits further growth of the core by removal of the resulting
 fragments due to radial drift caused by gas drag. However, the
 emergence of the pressure maximum halts the drift of the fragments,
 while their orbital eccentricities and inclinations are efficiently damped by
 gas drag.  As a result, the core of Saturn rapidly grows via accretion
 of the fragments near the pressure maximum.  We have found that in the
 minimum-mass solar nebula, kilometer sized planetesimals can produce a
 core exceeding 10 Earth masses within two million years.  Since Jupiter
 may not have undergone significant type II inward migration, it is
 likely that Jupiter's formation was completed when the local disk mass has
 already decayed to a value comparable to or less than Jovian mass.  The
 expected rapid growth of Saturn's core on a timescale comparable to
 or shorter than observationally inferred disk lifetime enables Saturn
 to acquire the current amount of envelope gas before the disk gas is
 completely depleted. The high heat energy release rate onto the core
 surface due to the rapid accretion of the fragments delays onset of
 runaway gas accretion until the core mass becomes somewhat larger than
 that of Jupiter, which is consistent with the estimate based on
 interior modeling. Therefore, the rapid formation of Saturn induced by
 gap opening of Jupiter can account for the formation of multiple gas
 giants (Jupiter and Saturn) without significant inward migration and
 larger core mass of Saturn than that of Jupiter.
\end{abstract}


\vspace{1cm}

\section{INTRODUCTION}
\label{sc:intro}

Since Jupiter resides at 5.2\,AU, it is likely that Jupiter did not
undergo significant type II migration \citep[e.g.,][]{ida_lin08}.  The
mass of the disk surrounding Jupiter decayed to a level comparable to or
less than Jupiter mass before Jupiter completed its formation and opened
up a gap. Since the disk could not push Jupiter, its type II migration
is negligible.  The formation of Saturn becomes problematic along this
line. With an assumption that cores grow through the collisional
accretion of surrounding planetesimals, the accretion timescale for
Saturn's core may be 5--10 times longer than that for Jupiter's
core. Therefore, since Jupiter forms on a timescale comparable to a disk
depletion timescale, a few Myrs, Saturn's core formed on a timescale of
10--30 Myrs. However, when the core formed, the disk mass should have
been so severely depleted that the core cannot accrete disk gas as
massive as the present envelope mass. To reconcile the inconsistency,
Saturn's core should have formed on a timescale shorter than several
Myrs after Jupiter's formation. 
\rev{
\revs{
Note that a scenario by \citet{walsh} where Jupiter first migrates
inward, and then outward 
\citep[e.g.,][]{morbidelli_crida}, 
also requires rapid formation of Saturn. }
}

\revs{A single planetary embryo is formed from collisional
coagulation of planetesimals in an annulus of the disk along the
embryo's orbit and further grows 
through collisions with surrounding remnant planetesimals.} An embryo
reaches the critical core mass, $\sim 10 M_\oplus$, to start gas
accretion for Saturn formation \citep{mizuno80,bodenheimer86,ikoma00}.
However, since massive embryos enlarge the random motion of
planetesimals, collisions between planetesimals are destructive.
Embryos also grow by the accretion of fragments that result from such
collisions. The eccentricities and inclinations of fragments are well
damped by gas drag, and fragment accretion accelerates embryo growth. On
the other hand, since such small fragments quickly drift inwards, the
final mass of an embryo is much smaller than the critical core mass to
start gas accretion \citep{kobayashi+10,kobayashi+11,ormel_kobayashi}.
However, Jupiter may have opened up a gap in a disk to truncate gas
inflow to Jupiter at the present Jupiter's mass, which forms a radial
pressure maximum near the edge of the gap. Radial drift of fragments is
stalled at around the pressure maximum \citep{adachi76}.  In addition,
type I migration of a core is also stalled there \citep{tanaka,masset}.
Since a core grows via fragment accretion without the loss of fragments
and the core itself, rapid formation of Saturn's core may be realized
under these conditions.

In order to investigate the rapid formation of Saturn core near the edge
of the gap, we perform simulations, which can derive accurate solutions
in both limits dominated by collisional fragmentation \citep{kobayashi10} or
coagulation \citep{kobayashi+10}. In Section \ref{sc:pm}, we investigate
the radial drift of bodies near the pressure maximum.  We model a disk for a
simulation and briefly explain the method of the simulation in Section
\ref{sc:model}. The core growth derived by the simulations is shown in
Section \ref{sc:result}. We discuss our findings in Section
\ref{sc:disc} and present our conclusion in Section \ref{sc:conc}.

\section{RADIAL DRIFT NEAR PRESSURE MAXIMA}
\label{sc:pm}

The velocity distinction between the gas rotational velocity $v_{\rm
gas}$ and the Keplarian velocity $v_{\rm k}$ is given by $v_{\rm
K}-v_{\rm gas} = \eta v_{\rm K}$ with \citep{adachi76}
\begin{eqnarray}
\eta &=& - \frac{1}{2\rho_{\rm gas} a \Omega_{\rm K}^2} \frac{d P_{\rm gas}}{d
  a},\label{eq:eta}  
\end{eqnarray}
where $\rho_{\rm gas}$ and $P_{\rm gas}$ are, respectively, the gas
density and pressure of disk midplane, $\Omega_{\rm K}$ is the
Keplarian angular velocity, and $a$ is the distance from the sun.
Equation~(\ref{eq:eta}) shows $\eta = 0$ at a pressure maximum.

Gap opening by Jupiter changes the radial profile of gas surface density,
which determines $\eta$. \citet{tanigawa} presented an analytical
formula for the gas surface density $\Sigma_{\rm g}$ around the gap
as\footnote{\revs{\citet{crida} derived a similar formula for the density
profile of gap. The locations of the pressure maximum (8.9\,AU for $\alpha =
6.9\times 10^{-5}$; 7.2\,AU for $\alpha = 10^{-3}$, where we adopt
$c/\Omega_{\rm K} a \approx 0.05$ at $a = a_{\rm J}$) are almost the same as
the values derived from the
formula by \citet{tanigawa}.}}
\begin{eqnarray}
 \Sigma_{\rm g} &=& \Sigma_{{\rm g}, \infty} \exp[ - (a-a_{\rm J}/l)^{-3}],\label{eq:sigma_gap}
\\
  l &=& \left[ \frac{8}{81 \pi} \frac{a_{\rm J}^2 \Omega_{\rm K}(a_{\rm
       J})}{\nu} \left(\frac{M_{\rm J}}{M_\sun}\right)^2\right]^{1/3}
  a_{\rm J}, 
\end{eqnarray}
where $a_{\rm J}$ is the orbital radius of Jupiter, the viscosity
coefficient $\nu$ is given by $\alpha c^2 \Omega_{\rm K}^{-1}$ with the
sound velocity $c$ and a parameter $\alpha$, $\Sigma_{{\rm g},\infty}$
is the gas surface density far away from Jupiter, and $M_{\rm J}$ and
$M_\sun$ are, respectively, the masses of Jupiter and the sun.  
We assume $\Sigma_{{\rm g},\infty} \propto a^{-p}$ and $c \propto a^{-q}$.
Equations~(\ref{eq:eta}) and (\ref{eq:sigma_gap}) then give $\eta$ near
the pressure maximum formed by Jupiter as
\begin{equation}
 \eta = \frac{c^2}{2 a^2 \Omega_{\rm K}^2} \left[ p + q +\frac{3}{2} -
					 \frac{3 l^3 a}{(a-a_{\rm
					 J})^4}\right].\label{eq:eta_gap} 
\end{equation}
\revs{Note that Equation~(\ref{eq:sigma_gap}) gives the density profile
determined by viscous torques. The thermal instability of the density profile
(Rayleigh criterion) regulates the density profile in the vicinity of
Jupiter \citep{ida_lin04,tanigawa,crida}. However, for the purpose of finding the location of the pressure
maximum, we can use Equation (\ref{eq:sigma_gap}) as long as
$c/\Omega_{\rm K} \la 0.1 a_{\rm J}$ and $\alpha \la 0.02$. 
}

Bodies lose their angular momenta through gas drag and then drift
inward, resulting in their radial transport. 
The gas drag force acting on a body with radius $r$ and mass $m$ for the relative velocity $u$ between the body and gas is expressed by $C_{\rm D} \pi r^2 \rho_{\rm gas}
u^2/2$ with the dimensionless gas
drag coefficient $C_{\rm D}$, which is determined taking into
account Epstein and Stokes regimes as well as constant $C_{\rm D}$ for 
high Reynolds number. 
The stopping time normalised by $\Omega_{\rm K}^{-1}$ is given by 
\begin{equation}
 \tilde \tau_{\rm stop} = a \Omega_{\rm K}^2 \tau / u,  
\end{equation}
where  
\begin{equation}
 \tau = \frac{2m}{\pi r^2 C_{\rm D} \rho_{\rm gas} v_{\rm K}}. 
\label{eq:tau_gas} 
\end{equation} 
\revs{
We do not consider very small fragments of $\tilde \tau_{\rm stop} \ll
1$ because radial drift of these fragments 
is negligible due to strong coupling with gas. }
The relative velocity $u$ then depends on eccentricity and inclination of a
body as well as on $\eta$. 
Eccentricities and inclinations of bodies have the Rayleigh-type
distributions with dispersions $e^*$ and $i^*$, respectively
\citep{ida92a}. 
The averaged value of $u$ is estimated as $(e^* + i^* + \eta) a \Omega_{\rm K}$.  
The drift velocity $v_{\rm d}$ is analytically given in the limit of $\tilde \tau_{\rm stop} \la 1$ or
$\tilde \tau_{\rm stop} \gg 1$ \citep{adachi76}. 
Averaging the solutions for bodies with eccentricities and
inclinations in Rayleigh-type distribution and connecting both these limits,
we then have
\footnote{\citet{inaba01} modified terms related to $\eta$ in the analytical formula for $\tilde{\tau}_{\rm stop} \gg
1$. We apply the modified formula for terms related to $\eta$ but the
cubic terms of $e^*$ and $i^*$ from \citet{adachi76}. } 
\begin{eqnarray}
 v_{\rm d} &=& - 2 \frac{a}{\tau} \frac{\tilde \tau_{\rm stop}^2}{1+
  \tilde \tau_{\rm stop}^2} 
  \left[ \eta (3.0 e^{*2} + 1.3 i^{*2} + \eta^2 )^{1/2 } \right. 
  \nonumber
\\
 && 
  \left. 
  +3.5 e^{*3} +0.42 i^{*3} 
  \right]. \label{eq:vd_final}
\end{eqnarray}
Equation~(\ref{eq:vd_final}) covers both limits in $\tilde \tau_{\rm stop} \la 1$ and
$\tilde \tau_{\rm stop} \gg 1$. 
Although the drift velocity becomes very low at the pressure
maximum because of $\eta = 0$, bodies with high $e^*$ or $i^*$ can drift
due to the cubic terms of $e^*$ and $i^*$. 

\begin{figure}[htbp]
\epsscale{1.} \plotone{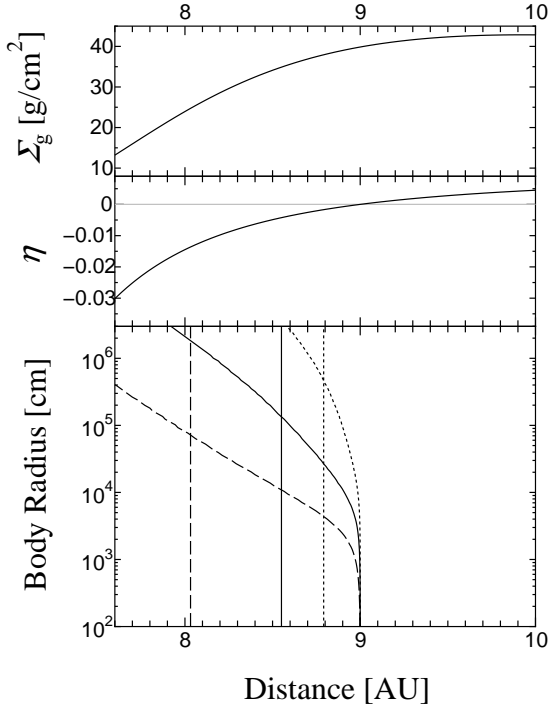} \figcaption{
The radial distribution of the surface density $\Sigma_{\rm g}$ (top) and
 $\eta$ given by Equation (\ref{eq:eta_gap}) (middle) and 
the location of $v_{\rm d} = 0$ vs. the radius of bodies (bottom). 
In order to determine the location, eccentricities of the bodies are given by the gas drag damping and the stirring of an embryo with 0.1 (dotted lines), 1 (solid line), 10
 (dashed lines) Earth masses. The vertical lines indicate the inner edge
 of the feeding
 zone of the embryo residing at 9\,AU.  
\label{fig:bodies_around}}
\end{figure}

A massive embryo produced in an annulus of the disk stirs surrounding
smaller bodies. \revs{The stirring and damping due to gas drag determine $e^*$
and $i^*$ of bodies, depending on their radii. }
Since $e^*$ and $i^*$ increase with
embryo growth, bodies drift inward from the pressure maximum. However,
since $\eta$ is negative inside the pressure maximum, the inward drift
due to $e^*$ and $i^*$ is canceled out at a certain distance by the
outward one caused by negative $\eta$.  For $\alpha = 6.9 \times
10^{-5}$, $p=3/2$, and $q = 1/4$, the radial distribution of
$\Sigma_{\rm g}$ and $\eta$ is shown in Fig.~\ref{fig:bodies_around}
\footnote{\revs{We discuss the applied $\alpha$ value below.}};
the pressure maximum is then located at 9\,AU.
Fig.~\ref{fig:bodies_around} also shows the location of $v_{\rm d} = 0$
as a function of the radius of bodies.  The location of $v_{\rm d} = 0$
for small bodies is close to 9\,AU, because $e^*$ and $i^*$ are strongly
damped by gas drag. Although the location of $v_{\rm d} = 0$ moves
inward for a massive embryo because of its strong stirring, the feeding
zone of the embryo, of which the width is given by 10 times Hill radii
of the embryo \citep{kokubo98}, is so wide that the embryo can accrete
kilometer-sized or smaller bodies. Although larger bodies drift out of
the feeding zone, the drift timescale is much longer than the timescale
of collisional fragmentation among them or of the accretion onto the
embryo. Note that even if such large bodies drift out of the feeding
zone, small bodies resulting from collisional fragmentation of the large
bodies drift outward and halt inside the feeding zone. Therefore, the
assumption of $v_{\rm d} \approx 0$ is valid to investigate embryo grow
at the pressure maximum.

The location of the pressure maximum depends on the turbulent strength,
$\alpha$ value.  Magneto-rotational instability may cause turbulence in
the disk; its activity depends on an ionization degree and the strength
of magnetic field. Because of the low ionization degree, a magnetically
decoupled midplane dead zone emerges at $\la 20\,$AU
\citep{sano}\footnote{\revs{Although a reduction of the vertical optical
depth of the disk due to dust depletion could make the dead zone small
\citep{sano}, in our model very small bodies supplied by collisional fragmentation, whose total mass is negligible as shown in
Section \ref{sc:result}, can produce a sufficient optical
depth.  } }. In the dead zone, the $\alpha$ value is expected to be as
low as $10^{-5}$--$10^{-4}$ if the net vertical magnetic field is weak
enough \citep{okuzumi11,gressel}; the pressure maximum is then located
at around 9\,AU.  For larger $\alpha \approx 10^{-3}$, the pressure
maximum moves to about 7\,AU. However, the behavior of bodies is
similar; bodies smaller than 1--10\,km halt inside the feeding zone of
an embryo formed at the pressure maximum and thereby these bodies
contribute to the embryo growth.  Therefore, the location of pressure
maximum (or choosing $\alpha$ value) insignificantly affects the Saturn
core formation. We thus fix the pressure maximum at 9\,AU in the
following.

\section{MODEL}
\label{sc:model}

We introduce a disk model for the initial surface mass density of solids $\Sigma_{\rm s,0}$ and gas $\Sigma_{\rm g,0}$ 
such that 
\begin{eqnarray}
 \Sigma_{\rm s,0} &=& x_{\rm s}\Sigma_{\rm MMSN,s} \left(\frac{a}{9{\rm AU}}\right)^{-3/2}
  \, {\rm g\, cm}^{-2}, \\
\Sigma_{\rm gas,0} &=& x_{\rm g} \Sigma_{\rm MMSN,g} \left(\frac{a}{9{\rm
					  AU}}\right)^{-3/2} 
  \, {\rm g\, cm}^{-2},\label{eq:sigma_gas} 
\end{eqnarray}
where $\Sigma_{\rm MMSN,s} = 1.1 \,{\rm g\,cm^{-2}}$ and $\Sigma_{\rm
MMSN,g} = 89 \,{\rm g\,cm^{-2}}$ are, respectively, the solid and gas
surface densities at 9\,AU in the minimum-mass solar nebula (MMSN) model
\citep{hayashi}.  We vary scaling factors $x_{\rm s}$ and $x_{\rm g}$ to
find the conditions for Saturn core formation.  Solid surface density
evolves in the simulation but gas density is set to be constant with
time.  
\revs{
We consider the disks with $x_{\rm g} \sim 1$ which are small enough for
Jupiter to reside at the current position without significant type II
migration and large enough to form cores for Jupiter and Saturn
\citep{ida_lin08}. 
}

We consider a disk from 8.6\,AU to 35\,AU.  The disk is divided into 14
annuli whose widths are given by 0.1 times the distance from the annulus
center to the sun.  The innermost annulus centered at $9{\rm \,AU}$ is
assumed to be the pressure maximum zone.  \rev{As shown in Section
\ref{sc:pm}, kilometer-sized or smaller bodies stay in the feeding zone
of an embryo residing at the pressure maximum.  In the MMSN disk, drift
time required for larger bodies to go out of the feeding zone of an
embryo with mass $\leq 10 M_\oplus$ is longer than the formation timescale of Saturn's
core, 1--$2 \times 10^{6}$ years, if the bodies are larger than about
30\,km.  For bodies with intermediate sizes, the drift time is longer
than their collisional fragmentation timescale \citep{kobayashi10}.
Therefore, we ignore radial drift of bodies in the innermost annulus
($v_{\rm d} = 0$).} The values of $\Sigma_{\rm g, 0}$ and $\eta$ are
taken from those at the radial center of each annulus, given by
Equation~(\ref{eq:sigma_gas}) and $\eta=5.4\times 10^{-3}(a/9{\rm
AU})^{1/2}$ except for the innermost annulus that has $\eta = 0$.  Each
annulus has a lot of mass bins to calculate the evolution of the mass
distribution of bodies via collisional coagulation and fragmentation
among bodies.  Collisional outcomes are assumed to be scaled by the
dimensionless impact energy $\phi = m_1 m_2 v^2/ 2 (m_1+m_2)^2 Q_{\rm
D}^*$ according to the simple model that \citet{kobayashi10} and
\citet{kobayashi+10} constructed consistently with laboratory
experiments and collisional simulations, where $m_1$ and $m_2$ are the
collider masses, $v$ is the collisional velocity between the colliders
and $Q_{\rm D}^*$ is the specific energy needed to eject half of the
total mass of the colliders.  The collisional cross sections of bodies
depend on their random velocities determined by $e^*$ and $i^*$.  Since
the stirring of large bodies increases the velocities, the mass
distribution of bodies affects the velocity evolution. Therefore, we
calculate the mass and velocity evolution simultaneously using the
collision rates \citep{inaba01} and the velocity evolution rates due to
mutual interaction of bodies \citep{ohtsuki02} and due to gas drag
\citep{adachi76}.  With a use of an upwind scheme, we calculate radial
transport of bodies due to radial drift given by
Equation~(\ref{eq:vd_final}) except for the innermost annulus.

In addition, we include the enhancement of collisions with embryos 
due to their tenuous
atmospheres and due to gas drag for small bodies with $\tilde
\tau_{\rm stop} \sim 1$, where $\tilde \tau_{\rm stop} = 1$ corresponds
to several tens centimeter in radius at 9\,AU. 
 We apply the atmospheric enhancement modeled
by \citet{inaba_ikoma03} with the grain depletion factor $f=1\times
10^{-2}$ and the collision rate between embryos and
small bodies with $\tilde \tau_{\rm stop} \sim 1$, given by
\citet{ormel_klahr}. For the Ormel \& Klahr model, we adopt the headwind
velocity of gas to an embryo as $(\eta + e_{\rm E} + i_{\rm E}) v_{\rm
K}$, where $e_{\rm E}$ and $i_{\rm E}$ are, respectively, eccentricity
and inclination of the embryo. 

Since small bodies are strongly coupled with gas flow, the model of
Ormel \& Klahr gives a very long timescale for collisions between very
small bodies and embryos.  In addition, because the collisional cascade
caused by embryo growth stalls at a body size of about 1--10\,m, the
total mass of smaller bodies is negligible \citep{kobayashi+10}.
Therefore, we neglect bodies with radius smaller than $r_{\rm min}$,
where $r_{\rm min} = 0.01$\,cm in our simulation. Indeed, embryo growth
is independent of $r_{\rm min}$ if $r_{\rm min} \ll 1$\,cm.

\section{SATURN CORE FORMATION}
\label{sc:result}

\begin{figure}[htbp]
\epsscale{1.} \plotone{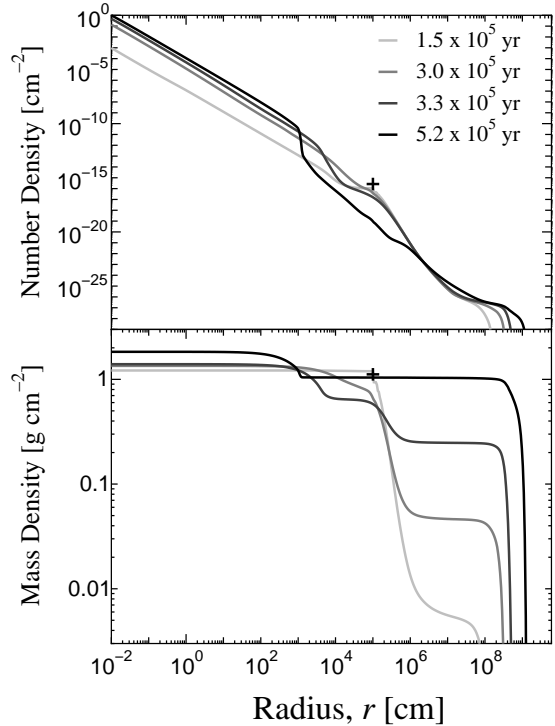} \figcaption{
Evolution of surface number and mass densities of bodies larger than
 radius $r$ at 9\,AU, starting from a swarm of
 planetesimals with initial radius $r_0 = 1$\,km in the MMSN model. Crosses
 indicate the initial condition and lines are size distributions at different times. 
\label{fig:zerovd_cumnum}}
\end{figure}

We perform simulations for core formation and growth starting from a
monodisperse mass population of planetesimals of mass $m_0$ and radius
$r_0$ with $e^* = 2 i^* = (2 m_0/3M_\odot)^{1/3}$. In
Fig.~\ref{fig:zerovd_cumnum}, the evolution of the size distribution
due to collisions is shown for $r_0 = 1\,$km ($m_0 = 4.2 \times
10^{15}$\,g) in the MMSN model ($x_{\rm s} = x_{\rm g} =
1$). Collisional coagulation of planetesimals produces a small number
of large bodies, planetary embryos. As embryos grow via the accretion
of a swarm of planetesimals, massive embryos stir planetesimals and
collisional fragmentation among planetesimals turns planetesimals into
small fragments.  Since gas drag suppresses the random motions of very
small bodies, fragmentation by collisions among such small bodies no
longer occurs.  The grinding down of bodies through collisions halts
at $\sim 10$\,m, where bodies accumulate (see
Fig.~\ref{fig:zerovd_cumnum}). The mass distribution is given by three
discrete components that are the largest bodies (planetary embryos),
initial sized or slightly larger bodies (planetesimals), and small
bodies with 1--100\,m.

The gas drag coefficient $C_{\rm D}$ is constant for kilometer-sized or
larger bodies but it increases with decreasing size in the Stokes
regime, given by $C_{\rm D} = 5.5 c l_{\rm g}/u r$
\citep{adachi76}. Here, $l_{\rm g} = l_{\rm g,0} / \rho_{\rm gas}$ is
the mean free path of gas molecular with $l_{\rm g,0} = 1.7\times
10^{-9} \,{\rm g\, cm}^{-2}$. The fragments accumulating at 1--100\,m
are controlled by the Stokes drag.  The equilibrium eccentricity $e_{\rm
eq}^*$ in the Stokes regime is estimated as
\citep{kobayashi+10,kobayashi+11}
\begin{equation}
 e_{\rm eq}^{*2} = \frac{h_M^3 \langle P_{\rm VS} \rangle \tau a \Omega_{\rm K}^2 }
  {2^{7/3} \pi \tilde{b} u},\label{eq:eeq} 
\end{equation}
where $h_M = (M/3 M_\sun)^{1/3}$ is the reduced Hill radius of an embryo
with mass $M$, $\langle P_{\rm VS} \rangle = 73$ is the dimensionless
stirring rate for $e^* \ll h_M$, and $\tilde{b} = 10$ is the separation
of embryos divided by their mutual Hill radius. Since $\tau \propto u$
in the Stokes regime, $e_{\rm eq}^*$ is independent of $u$; the
eccentricity is the same as that in the case $\eta \neq 0$.  When a body
of mass $m$ collides with a similar-sized body, the impact energy is
given by $m (e_{\rm eq}^* v_{\rm k})^2/4$. If the energy is much smaller
than $2 m Q_{\rm D}^*$, the collision no longer produces significant
fragments. The collisional cascade stalls around $(e_{\rm eq}^* v_{\rm
k})^2 = C_{\rm L} Q_{\rm D}^*$ with a constant $C_{\rm L} \sim 1$ and
small bodies accumulate around the radius satisfying the condition,
which is obtained from the condition and Equation (\ref{eq:eeq}) as
\begin{eqnarray}
 r_{\rm f} &=& 7.9 \, C_{\rm L}^{1/2} 
\left(
\frac{Q_{\rm D}^*}{8.8 \times 10^6 {\rm \, erg\, g}^{-1}} 
\right)^{1/2} 
\nonumber
\\
&&\times 
\left(
\frac{T}{93\,{\rm K}}
\right)^{1/4}
\left(
\frac{a}{9\,{\rm AU}}
\right)^{5/4}
\left(
\frac{M}{M_\oplus}
\right)^{-1/2}
\,{\rm m},\label{eq:rf} 
\end{eqnarray}
where $T$ is the disk midplane temperature and the value of $Q_{\rm
D}^*$ for a body of radius $r=1\,{\rm m}$ is applied \citep{benz99}.
Indeed, $r_{\rm f}$ is consistent with the result of simulations for
$C_{\rm L} \sim 1$ (see Fig.~\ref{fig:zerovd_cumnum}). Interestingly,
$r_{\rm f}$ is independent of $\eta$ and $\rho_{\rm gas}$ because gas
drag is determined in the Stokes regime.  In addition, $e^*_{\rm f} =
(C_{\rm L} Q_{\rm D}^*)^{1/2}/v_{\rm K} \approx 3 \times 10^{-3}
(a/9\,{\rm AU})^{1/2}$ is smaller than $h_M = 10^{-2}
(M/M_\oplus)^{1/3}$, resulting in a very large accretion rate of the
fragments onto an embryo \citep{ida89}.

\begin{figure}[htbp]
\epsscale{1.} \plotone{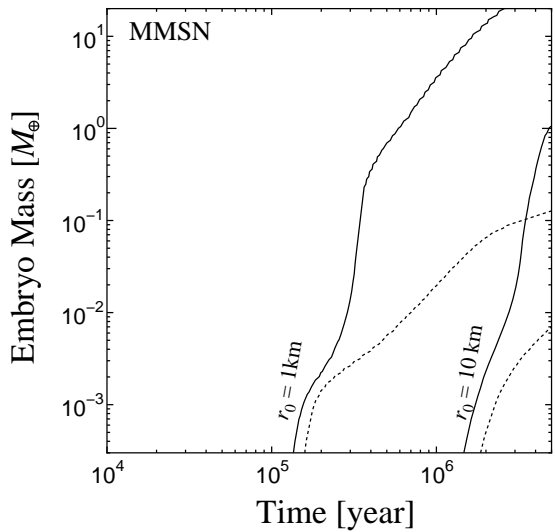} \figcaption{
Embryo growth at 9\,AU for $r_0 = 1$\,km and 10\,km in the MMSN
 model ($x_{\rm g} = x_{\rm s} = 1$). 
\revs{For reference, dotted lines show the results in the case
 of an unperturbed disk (i.e. without Jupiter's gap)}. 
\label{fig:zerovd_s1}}
\end{figure}

Fig. \ref{fig:zerovd_s1} shows the evolution of embryo
masses\footnote{In our simulation, we define ``runaway bodies'' which
cannot collide with each other due to their large orbital separations
\citep[see][]{kobayashi+10}. Here ``embryo masses'' represent the
averaged mass of runaway bodies, which roughly corresponds to the mass
of the largest body.} at 9\,AU for the MMSN disk with $r_0 = 1$\,km and
10\,km. Embryos initially grow exponentially with time and the growth
time scale is proportional to $r_0 / \Sigma_{\rm s,0}$
\citep{ormel10a,ormel10b}, which is called runaway growth. Massive
embryos formed by the runaway growth stir remnant planetesimals and then
slower, oligarchic growth starts for embryos larger than $\sim 10^{-3}
M_\oplus$.  Such massive embryos then activate collisional fragmentation
of planetesimals.  The very low $e^*$ and $i^*$ of the small resulting
fragments damped by strong gas drag tend to accelerate the embryo
growth.  On the other hand, the gas drag also leads to the radial drift
of the fragments, which generally limits the embryo growth
\citep{kobayashi+10,kobayashi+11}.  However, near the pressure maximum
due to gap opening by Jupiter, embryos effectively accrete such
fragments without the loss of the fragments. In
Fig.~\ref{fig:zerovd_s1}, the embryo growth starts being accelerated at
$\sim 0.01 M_\oplus$ (at about $3\times10^5$ years for $r_0 = 1\,$km and
at about $3\times 10^6$ years for $r_0 = 10$\,km), because fragment
production becomes efficient at such an embryo mass. At $\ga M_\oplus$,
the acceleration slows down due to depletion of the total amount of
surrounding bodies by the embryo accretion.  Nevertheless, the embryo
keeps growing beyond the ``isolation mass'', which is several Earth
masses for the MMSN model, because fragments produced in outer annuli
drift into the region at $\sim 9\,$AU.  Eventually, an embryo reaches 10
Earth masses in $2\times 10^6$ years for $r_0 = 1\,$km, while the embryo
mass is about an Earth mass in $5 \times 10^6$ years for $r_0 = 10$\,km.

\rev{In the oligarchic growth of embryos, embryos have orbital
separations of 10 Hill radii of themselves \citep{kokubo98}. 
Since 10 Hill radii of an embryo with $10M_\oplus$ at 9\,AU, about 2\,AU, is larger than the pressure
maximum region (see Fig.~\ref{fig:bodies_around}), 
only an embryo larger than $10M_\oplus$ can be formed in the pressure maximum. 
In case that the pressure maximum does not emerge (dotted lines
in Fig.~\ref{fig:zerovd_s1}), embryo growth is not so rapid and stalls
at about $0.1 M_{\oplus}$, 
because surrounding
planetesimals are lost due to combination of planetesimal fragmentation
with the radial drift of resulting fragments \citep[see also][]{kobayashi+10,kobayashi+11}. 
Since there is no pressure maximum in the outer disk, 
embryos beyond 10\,AU cannot become larger than a body of about Mars mass. 
Meanwhile the fragmentation of planetesimals and the ensuing radial drift of
resultant fragments remove the solid mass reservoir in the outer disk, 
\revs{
which facilitates the formation of a large enough embryo in the
pressure-maximum region. 
}
}

\begin{figure}[htbp]
\epsscale{1.} \plotone{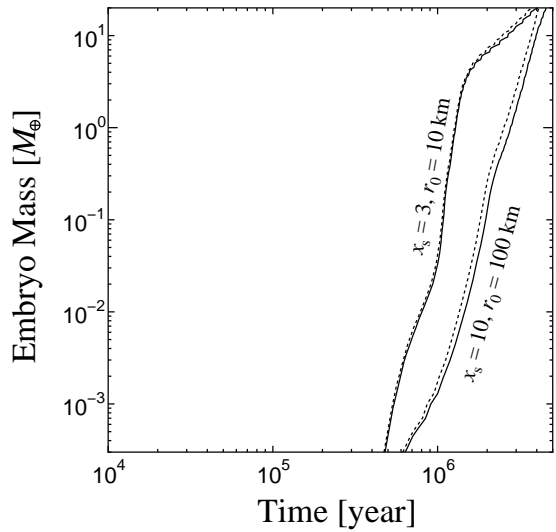} \figcaption{
Embryos can reach 10 Earth masses at 9\,AU in a massive disk with $x_{\rm s} = 3$ for $r_0 = 10$\,km
 and with $x_{\rm s}= 10$ for $r_0 = 100$\,km. 
The gas density of the disk is determined as $x_{\rm g} = 1$ (solid lines) and
 $x_{\rm g} = x_{\rm s}$ (dotted lines). 
\label{fig:zerovd_mix}}
\end{figure}

\revs{ 
Although a core can keep growing at the pressure maximum, 
the core is required to reach $\sim 10 M_{\oplus}$ within several Myrs 
for the formation of Saturn. 
}
For
$r_0 \ga 10\,$km, an MMSN disk ($x_{\rm s} = x_{\rm g} = 1$) cannot
produce a core with 10 Earth masses in several million years (see
Fig.~\ref{fig:zerovd_s1}). Fragment production from large planetesimals
is ineffective because of large $Q_{\rm D}^*$ and occasional collisions
among them, which delays embryo growth via fragment accretion.  A large
solid surface density ($x_{\rm s} > 1$) that increases fragment
production is necessary for planetesimals with larger $r_0$ to form a
massive core in several million years. A core can reach 10 Earth masses
in a disk with $x_{\rm s} \ga 3$ for $r_0=10$\,km and with $x_{\rm s}
\ga 10$ for $r_0=100$\,km (see Fig.~\ref{fig:zerovd_mix}).  Note that
the growth of cores is almost independent of $x_{\rm g}$, because
fragments with $r_{\rm f}$ contributing most to the growth are
controlled by Stokes drag \citep[e.g.,][]{beauge}.

\section{DISCUSSION}
\label{sc:disc}

In this paper, we have taken the size of initial planetesimals as a free
parameter. The size depends on the formation mechanisms of
planetesimals. If planetesimals are formed via self-gravitational
instability in a thin dust layer
\citep{goldreich_ward,michikoshi,takeuchi} or of dust accumulated in the
structure of disk turbulence \citep{johansen07,cuzzi08}, resulting
planetesimals are 10\,km or larger. On the other hand, collisional
coagulation of fluffy dust is possible to produce planetesimals
overcoming an obstacle of rapid radial drift at $\tilde \tau_{\rm stop}
\sim 1$ \citep{okuzumi}; the resulting planetesimals may be smaller.  In
addition, bodies in a significant amount drift from the outer disk
during the collisional growth of fluffy dust and such drifting dust
increases the solid surface density form MMSN to about $4\times$MMSN
inside $\sim 10$\,AU \citep{okuzumi}.  Therefore, we should address the
history of Saturn core formation including planetesimal formation.

\rev{A planetary embryo born at the pressure maximum migrates due to
interaction with the surrounding gas disk (type I).  The time required
to go through half the width of the embryo's feeding zone due to type I
migration is longer than the growth time of the embryo unless the embryo
exceeds $0.4M_\oplus$ \citep{tanaka}.  
The type I migration rate of an embryo with mass $M$ at distance $a$ is
given by $da/dt = 2 f_{\rm I} M \Sigma_{\rm g} a^5 \Omega_{\rm K}^3 /
M_\sun^2 c^2$, where $f_{\rm I}$ is a factor that is determined by $p = - d \ln
\Sigma_{\rm g} / d \ln a$  and $q = - d \ln c / d \ln a$. 
If horseshoe torques are fully saturated, 
$f_{\rm I} = -1.45 - 0.6 p - 1.8 q$ in a locally isothermal disk and
$f_{\rm I} = -1.04 - 0.77 p - 0.77 q$ in an adiabatic disk
\citep{paardekooper11}. In the outer disk where we assume $p = 1.5$ and
$q = 0.25$, type I migration is inward due to a negative value of
$f_{\rm I}$, while outward migration occurs around the gap opened by
Jupiter where the radial $\Sigma_{\rm g}$ slope has a large positive
value. Therefore type I migration stalls 
at a certain distance under the density profile shown in
Fig.~\ref{fig:bodies_around}; 
the location of
zero migration is at 8.6\,AU in the locally isothermal disk
and at 9.1\,AU in the adiabatic disk. 
Since the location is near the pressure maximum, such a large embryo,
which has a wide feeding zone, can grow via the accretion of bodies
accumulating at around the pressure maximum.  Note that if the torques
are fully unsaturated, $f_{\rm I} = -0.85 - p + 1.8 q$ in the locally
isothermal disk and $f_{\rm I} = -0.61 -2.33 p + 2.82 q$ in the adiabatic
disk \citep{paardekooper10}. 
\revs{
Then, $f_{\rm I}$  is positive (outward migration) if $q$ has a large
value. In the optically thick regions (inner disks or massive disks)
where viscous heating is more dominant than irradiation, $q$ has a
relatively large value. However, in the outer regions ($\sim 9\,$AU) of
moderate disks with $x_{\rm g} \sim 1$ that we are concerned with, it is
likely that the disk temperature is determined by irradiation rather
than viscous heating and that $q$ has a relatively small value ($\sim
0.25$). Then, for $p = 1$--$1.5$, type I migration is inward ($f_{\rm I}
< 0$) and the argument of the migration trap is still applicable. 
Furthermore, the torque saturation is likely in the disk of interest. 
The saturation is prevented only if the horseshoe liberation time is
comparable to the diffusion/thermal diffusion timescale across the
horseshoe width \citep{paardekooper11} and if an embryo has a low enough
eccentricity \citep{bitsch}.  The embryos resulting form our simulations
have eccentricities larger than 0.01. Therefore the condition for type
I migration in the outer disk where Saturn's core forms is most likely
to result in inward migration, except in the vicinity of the
gap. Consequently, type I migration does not inhibit the fast formation
of Saturn's core but rather strengthens it.
}
}

Once a core exceeds a critical core mass, the core has no longer a
static atmosphere and rapid gas accretion starts to form a gas giant.
Interior modeling of Jupiter and Saturn suggests that Jupiter's core
mass is smaller than $10M_\oplus$, while Saturn's core is larger than $10 M_\oplus$ \citep{guillot}.  Since heating by the
accretion of small bodies onto a core stabilises the atmosphere, the
critical core mass becomes large for a fast growth core
\citep[e.g.,][]{ikoma00}.  Traditional scenarios, in which the core
growth is 5--10 times slower for Saturn than for Jupiter, cannot explain
the estimate of core masses of Jupiter and Saturn.  However, Saturn core
grows rapidly at the pressure maximum caused by Jupiter's gap
opening. Since the growth is possible to be faster for Saturn
than for Jupiter, the resulting core masses of Jupiter and Saturn may be
consistent with those derived form the interior modeling.

Since the core of Saturn is estimated to be larger than $10\,M_\oplus$
\citep{guillot}, Saturn should be formed via core accretion.  We have
shown that if Jupiter has already been completed and opened a gap in
the disk, the massive core of Saturn has rapidly formed from
kilometer sized planetesimals in an MMSN disk or from larger
planetesimals in a disk with a larger amount of planetesimals.  On the other
hand, since small initial planetesimals are quickly removed by combination of
fragmentation with gas drag, the core of Jupiter is likely to
have formed from large initial planetesimals with $\sim 100\,$km at around
5\,AU in a $10\times$MMSN disk \citep{kobayashi+11}. Jupiter is possible to
form earlier than Saturn core formation, if either (1) the solar nebula
is as massive as self-gravitational instability occurs, (2) pressure
maxima originally exist in the solar nebula, or (3) 
\rev{planetesimals of size 100\,km are born around 5\,AU in a massive disk}. All the
cases are consistent with the sequential formation of Saturn.

After the gas accretion, Saturn might open up another gap in the disk,
resulting in the emergence of an additional pressure maximum. Neptune
and Uranus mainly consist of ice with more than $10M_\oplus$ covered
with tenuous atmospheres. Near the pressure maximum, such massive icy
planets can be formed. Fragments produced in the outer disk accumulate
within about the width of the disk scale height near the pressure
maximum but the feeding zone of such a massive planet is wider than the
scale height; hence only one planet can be produced. Meanwhile, the
orbital eccentricity distribution of Kuiper belt objects in the 3:2
mean-motion resonance with Neptune suggests that Neptune migrated
outward from around 20\,AU to the current position \citep{malhotra}.
Neptune was therefore expected to have formed at the pressure maximum
and to have then migrated outward due to interaction with the outer
planetesimal disk \rev{\citep[e.g.,][]{ida00,kirsh}.  Note that torques
from interaction with gas that cause type I migration suppress the planetesimal-driven
migration.  Since Saturn should form in gas with a high density, high
gas-to-solid surface densities render type I migration stronger
compared to the planetesimal-driven migration. Therefore, the
planetesimal-driven migration is negligible for the Saturn
formation. However, Neptune plausibly migrated outward through
interactions with planetesimals because of a low gas density.  Uranus
might then have formed at around the pressure maximum caused by
Saturn}, where fragments produced from remnant planetesimals that induce
the Neptune migration accumulate.  Therefore, we should also address
Uranus and Neptune formation after Saturn's formation.

\section{Conclusion}
\label{sc:conc}

We have investigated the core formation of Saturn at the pressure
maximum caused by Jupiter's gap opening in the solar nebula. Small
particles feel strong gas drag and drift radially. The loss of bodies
due to the drift stalls embryo growth before embryos reach the
critical core mass \citep{kobayashi+10,kobayashi+11}.  Although bodies
stirred by a large embryo can drift from the pressure maximum, the
drift halts near the pressure maximum due to positive radial slope of
gas pressure; hence an embryo at the pressure maximum effectively
grows without loss of surrounding bodies (see Section \ref{sc:pm}).
Starting from monodisperse planetesimals, planetary embryos are
generated via collisional evolution and stir remnant planetesimals,
resulting in fragmentation of planetesimals.  Fragments collide with
each other and they accumulate at a radius $r_{\rm f}$, given by
Equation~(\ref{eq:rf}). The random velocity of bodies with $r_{\rm f}$
is well damped by gas drag; thereby embryos rapidly accrete such
bodies. Fragments produced in the outer disk move to the pressure
maximum, which contribute to further embryo growth. Due to the rapid
accretion, a core as massive as 10 Earth masses forms in several
million years, for kilometer-sized initial planetesimals in an MMSN
disk, while the core formation needs a disk with 3 times larger solid
surface density for 10\,km planetesimals and with 10 times solid
surface density for 100\,km planetesimals.  The growth is almost
independent of gas density.  Since the rapid formation in a disk with
moderate mass is consistent with insignificant type II migration of
Jupiter and larger core of Saturn than that of Jupiter, Saturn's core
may have formed in the pressure maximum after Jupiter opened up a gap
in the solar nebula.

\vspace{1cm}

We thank S. Inutsuka for his comments and encouragement and H. Tanaka
and S. Okuzumi for useful discussion. We also acknowledge
D. Lin and E. Kokubo for inspiring us to consider the calculations of induced
formation model of Saturn. 
\rev{
For C.W.O. support for this work was provided by NASA through Hubble
Fellowship grant \#HST-HF-51294.01-A awarded by the Space Telescope
Science Institute, which is operated by the Association of Universities
for Research in Astronomy, Inc., for NASA, under contract NAS 5-26555. 
}

\end{document}